\documentclass[apj]{emulateapj}
\shorttitle{Shock Acceleration}
\usepackage{color}

\begin{document}
\newcommand{\myemail}{guofan@lpl.arizona.edu}
\newcommand{\joeemail}{giacalon@lpl.arizona.edu}
\newcommand{\randyemail}{jokipii@lpl.arizona.edu}
\newcommand{\jozsefemail}{kota@lpl.arizona.edu}
\renewcommand{\thefootnote}{\fnsymbol{footnote}}
\title{Particle acceleration by collisionless shocks containing large-scale magnetic-field variations}

\author{F. Guo\footnote{\myemail} ,\ J. R. Jokipii\footnote{\randyemail} \ and J. Kota\footnote{\jozsefemail}}
\affil{Department of Planetary Sciences and Lunar and Planetary Laboratory,
University of Arizona,
\\1629 E. University Blvd. Tucson AZ 85721}

\begin{abstract}
Diffusive shock acceleration at collisionless shocks is thought to be the
source of many of the energetic particles observed in space. Large-scale
spatial variations of the magnetic field has been shown to be important in
understanding observations. The effects are complex, so here we consider a
simple, illustrative model. Here, we solve numerically the Parker transport
equation for a shock in the presence of large-scale sinusoidal magnetic-field
variations. We demonstrate that the familiar planar-shock results can be
significantly altered as a consequence of large-scale, meandering magnetic
lines of force. Because perpendicular diffusion coefficient $\kappa_\perp$ is
generally much smaller than parallel diffusion coefficient $\kappa_\parallel$,
the energetic charged particles are trapped and preferentially accelerated
along the shock front in the regions where the connection points of magnetic
field lines intersecting the shock surface converge, and thus create the ``hot
spots" of the accelerated particles. For the regions where the connection
points separate from each other, the acceleration to high energies will be
suppressed. Further, the particles diffuse away from the ``hot spot" regions
and modify the spectra of downstream particle distribution. These features are
qualitatively similar to the recent Voyager's observation in the Heliosheath.
These results are potentially important for particle acceleration at shocks
propagating in turbulent magnetized plasmas as well as those which contain
large-scale nonplanar structures. Examples include anomalous cosmic rays
accelerated by the solar wind termination shock, energetic particles observed
in propagating heliospheric shocks, and galactic cosmic rays accelerated by
supernova blast waves, etc.
\end{abstract}

\keywords{acceleration of particles - cosmic rays - shock waves - magnetic
field}

\section{Introduction}

Collisionless shocks in space and other astrophysical environments are
efficient accelerators of energetic charged-particles. Diffusive shock
acceleration \citep[hereinafter
DSA;][]{Krymsky1977,Axford1977,Bell1978a,Blandford1978}, is the most popular
theory for charged-particle acceleration. It naturally predicts a universal
power-law distribution $f \varpropto p^{-\gamma}$ with $\gamma \sim 4.0$ for
strong shocks, where $f$ is the phase-space distribution function, close to
what observed in cosmic rays in many different regions of space. The basic
conclusions of DSA can be drawn from the Parker transport equation
\citep{Parker1965} by considering the shock to be a compressive discontinuity
in an infinite one-dimensional and time steady system. DSA is thought to be the
mechanism that accelerates anomalous cosmic rays (ACRs) in the Heliospheric
termination shock and also galactic cosmic rays (GCRs) with energy up to at
least $10^{15}$ eV in supernova blast waves. However, recent {\it in situ}
observations in the termination shock and the Heliosheath by \emph{Voyager} $1$
\citep{Stone2005} found the intensity of ACRs is not peaked at the termination
shock and the energy spectrum is still unfolding after entering the
Heliosheath, which strongly indicates the simple planar shock model is
inadequate to interpret the acceleration of ACRs. Numerical and analytical
studies suggest the possible solution can be made by considering the temporary
and/or spatial variation
\citep{Florinski2006GeoRL,McComas2006,Jokipii2008AIP,Kota2008AIP,Schwadron2008ApJ}.
In particular, \citet{McComas2006} discussed the importance of the magnetic
geometry of a blunt shock on particle acceleration. They argued that the
missing ACRs at the nose of the Heliospheric termination shock is due to
particle energization occuring primarily back along the flanks of the shock
where magnetic field lines have had a longer connection time and higher
injection efficiency. \citet{Kota2008AIP} presented a more sophisticated
simulation which gives results similar to that described by
\citet{McComas2006}. \citet{Schwadron2008ApJ} also developed a 3-D analytic
model for particle acceleration in a blunt shock, including perpendicular
diffusion and drift motion due to large-scale shock structure.

Large-scale magnetic field line meandering is ubiquitous in the heliosphere and
other astrophysical environments \citep{Jokipii1966,Jokipii1969,Parker1979}.
The acceleration of charged-particles in collisionless shocks has been shown to
be strongly affected by magnetic-field turbulence at different scales
\citep{Giacalone2005a,Giacalone2005b,Giacalone2008,Guo2010}. The large-scale
magnetic field variation will have important effects on the shock acceleration
since the transport of charged particles is different in the direction parallel
and perpendicular to the magnetic field, as shown in early work
\citep{Jokipii1982ApJ,Jokipii1987ApJ}. The blunt shocks and shocks with
fluctuating front \citep{Li2006AIP} which have the similar geometry, are also
relevant to this problem. In this study we consider the effect of the
large-scale spatial variation of magnetic field on DSA.

\section{Basic Considerations and Numerical Model}
The diffusive shock acceleration (DSA) can be studied by solving the Parker
transport equation \citep{Parker1965}, which describes the evolution of the
quasi-isotropic distribution function $f(x_i, p, t)$ of energetic particles
with momentum $p$ dependent on the position $x_i$ and time $t$ including
effects of diffusion, convection, drift, acceleration and source particles:

\begin{equation}
\frac{\partial f}{\partial t} = \frac{\partial}{\partial
  x_i}\left[\kappa_{ij}\frac{\partial f}{\partial x_j}\right] -
  U_{i}\frac{\partial f}{\partial x_i}-V_{d,i}\frac{\partial f}{\partial x_i}+
  \frac{1}{3}\frac{\partial U_{i}}{\partial x_i} \left[\frac{\partial f}{\partial \ln p} \right] +
  Q
\end{equation}

Here $\kappa_{ij}$ is the diffusion coefficient tensor, $U_i$ is the convection
velocity and $Q$ is a local source. The drift velocity is given by
$\textbf{V}_d = (pcw/3q)\nabla\times(\textbf{B}/B^2)$, where $w$ is the
velocity of the particle, $c$ is the speed of light, and $q$ is the electric
charge of the particle.

\citet{Kota2008AIP} and \citet[][manuscript to be submitted]{Kota2010}
considered analytically a model in which the upstream magnetic field was a
plane (say, $x, y$), with average direction in the $y$ direction. The
x-component of the magnetic field was composed of uniform sections (straight
field lines) alternating in sign, which were periodic in $y$. They find ``hot
spots" and spectral effects which illustrate the effect of an upstream
meandering in the magnetic field.

Here, we consider a 2-D $(x,z)$ system with a planar shock at $x=0$, and a
sinusoid magnetic field $\textbf{B} = \textbf{B}_0 + \sin(k z)\delta
\textbf{B}$. For most of the parts in this paper we discuss the case shown in
Figure 1. In this figure, the magnetic lines of force are illustrated by blue
lines. The shock is denoted by the red dashed line. The system is periodic in
the $z$ direction, with the magnetic field convecting from upstream ($x<0$) to
downstream ($x>0$). In the shock frame, the particles will be subjected to
convection and diffusion due to the flow velocities $U_1$(upstream) and
$U_2$(downstream) and diffusion coefficients parallel and perpendicular to
large scale magnetic field $(\kappa_{1(\parallel, \perp)} $\space and \space$
\kappa_{2(\parallel,\perp)})$, respectively. The gradient and curvature drifts
in this case are only in the direction out of the $x-z$ plane and thus
irrelevant to this study. Because of the steady velocity difference between
upstream and downstream, charged particles which travel through the shock layer
will be accelerated. However, since we consider the large-scale magnetic field
variation, transport of energetic particles in the fluctuating magnetic field
become important. The diffusion coefficient in the $x-z$ system can be
expressed as:

\begin{equation}
\kappa_{ij} = \kappa_\perp \delta_{ij} -
\frac{(\kappa_\perp-\kappa_\parallel)B_iB_j}{B^2}
\end{equation}

The normalization units chosen in this study are: the spatial scale $X_0 = 10$
AU, the upstream velocity $U_1 = 500$ km/s, the time scale $T_0 = 3\times 10^6$
sec and the diffusion coefficients are in unit of $\kappa_0 = 7.5 \times
10^{21} cm^2/s$. The shock compression ratio $r = U_1/U_2$ is taken to be
$4.0$. The shock layer is considered to be a sharp variation $U_x = (U_1+U_2)/2
- (U_1-U_2)tanh(x/th)/2$ with thickness $th = 1 \times 10^{-3}$, which is
required to be less than $\kappa_{xx1}/U_1$ everywhere in the upstream
simulation domain, where $\kappa_{xx1}$ is the upstream diffusion coefficient
normal to the shock surface. The simulation domain is taken to be $[-2.0<x<2.0,
-\pi<z<\pi ]$. The parallel diffusion coefficients upstream and downstream are
assumed to be the same and taken to be $\kappa_{\parallel1} =
\kappa_{\parallel2} = 0.1$ at $p = p_0$. The ratio between parallel diffusion
coefficient and perpendicular diffusion coefficient is taken to be
$\kappa_{\perp}/\kappa_{\parallel} = 0.05$, which is consistent with that
determined by integrating the trajectories of test particles in magnetic
turbulence models \citep{Giacalone1999}. The momentum dependence of the
diffusion coefficient is taken to be $\kappa \propto p^{4/3}$, corresponding to
non-relativistic particles in a Kolmogorov turbulence spectrum
\citep{Jokipii1971}. The time step is $1 \times 10^{-7}$, which is small enough
to resolve the variation of $U_x$ at the finite shock layer. We use a
stochastic integration method, which is described in the Appendix, to obtain
the numerical solution of the transport equation. The pseudo-particles are
injected at the shock with initial momentum $p_0$, and will be accelerated if
they across the shock. Particles which move past the upstream or downstream
boundaries will be removed from the simulation. The system is periodic in the
$z-$direction, so a pseudo-particle crossing the boundaries in $z$ will
re-appear at the opposite boundary and continue to be followed. A particle
splitting technique similar to \citep{Giacalone2005a} is used in order to
improve the statistics. Although we use very approximate parameters, we note
that the results are insensitive to the precise numbers.

\section{Results and Discussion}
\subsection{A shock propagating perpendicular to the average magnetic field}
Consider first the case where the average magnetic field is in the
$z$-direction and the fluctuating magnetic field is $\delta B = B_0$. As shown
in Figure 1, the magnetic field is convected through the shock front and is
compressed in the $x$ direction, thus $B_{z2} = rB_{z1}$. For the sinusoid
magnetic field considered in this paper, the local angle between upstream
magnetic field and shock normal, $\theta_{Bn}$, will vary along the shock
surface. As a magnetic field line passes through the shock surface, its
connection points (the points where the field lines intersect the surface of
the shock) will be moving apart in the middle of the plane ($z = 0$) and
approaching each other on the both sides of the system ($|z| = \pi$). Since,
$\kappa_\parallel \gg \kappa_\perp$, the particles tend to remain on the
magnetic field lines. Because the acceleration only occurs at the shock front,
as the magnetic lines of force convect downstream, the particles will be
trapped and accelerated at places where the connection points converge toward
each other, leading to further acceleration. For the regions where the field
lines separate from each other, the particles are swept away from those
regions. Figure $2$ displays the spatial distribution contours of accelerated
particles in three energy ranges: $3.0<p/p_0<4.0$ (top), $8.0<p/p_0<10.0$
(middle), and $15.0<p/p_0<30.0$ (bottom). The density is represented by the
number of particles in simulation and its unit is arbitrary. It can be seen
that ``hot spots" form in the regions that connection points approaching each
other at all energy ranges, with lobes extend along the magnetic field lines.
The density of the accelerated particles at the connection-point separating
region (in the middle of the plane) is clearly much smaller, although there is
still a concentration of low-energy accelerated particles there since the
acceleration of low-energy particles is rapid and efficient at perpendicular
shocks. At higher energy ranges (middle and bottom), the lack of accelerated
particles may be interpreted as due to the fact that the acceleration to high
energies takes time.

Figure 3 illustrates the profiles of the density of accelerated particles for
different energy ranges at $z = 0 $ (top) and $z=\pi $ (bottom). In each panel,
the black solid lines show the density of low energy particles
($3.0<p/p_0<4.0$), the blue dashed lines show the density of intermediate
energy particles ($8.0<p/p_0<10.0$), and red dot dashed lines show the density
of particles with high energies ($15.0<p/p_0<30.0$). In connection-point
separating regions $z = 0 $ (top), it can be seen that while the downstream
distribution of low energy particles is roughly a constant, the density of
particles with higher energies increase as a function of distance downstream
from the shock. These particles are not accelerated at the shock layer in the
center of the plane but in the ``hot spots". At $z = \pi $ (bottom), the
density of particles of all energies decreases as a function of distance, which
indicates the accelerated particles diffusive away from the ``hot spots". Since
the high energy particles have larger diffusion coefficients than the particles
with low energy, it is easier for them to transport to the middle of the plane.
The profile at $z = 0$ is similar to \emph{Voyager}'s observation of anomalous
cosmic rays (ACRs) at the termination shock and the Heliosheath
\citep{Stone2005,Cummings2008} which shows the intensity of the ACRs is still
increasing and the energy spectrum is unfolding over a large distance after
entering the Heliosheath. The same physics has discussed by
\citet{Jokipii2008AIP}, where ``hot spot" of energetic particles is produced by
the spatial variation of the injection of the source particles. In our current
work, the concentration of energetic particles are a consequence of particle
accelerated in a shock containing large-scale magnetic variation.

The top panel in Figure $4$ represents the positions in $z$ direction and the
times as soon as the particles reached a certain momentum $p_c = 3p_0$. We show
that particles are accelerated mainly at the connection-point converging
region. There are also particles accelerated at middle of the plane because the
particles can gain energy rapidly at perpendicular or highly oblique shock due
to the smallness of perpendicular diffusion coefficient \citep{Jokipii1987ApJ},
however, the further acceleration is suppressed by the effect that the charged
particles travel away from the connection-point separating region, see also the
top panel in Figure 5. It is clear that since the particles tend to follow the
magnetic field lines, when the field line connection points separate from each
other as field convects through the shock, the particles travel mainly along
magnetic field and away from the middle of the plane. The characteristic time
for a field line convect from upstream to downstream $\tau_c = D/U_1 \sim 1$,
therefore there is no significant acceleration in the middle of the plane after
$t = \tau_c$. Some of the particles can get more acceleration traveling from
other region to ``hot spots". Figure 4 $bottom$ shows the distance particles
traveled in $z$ direction from its original places $|z-z_0|$ versus time when
particle get accelerated at a certain energy ($p_c = 3p_0$). It shows many
particles are accelerated close to their original position, which is related to
the acceleration in the ``hot spot". Nevertheless, there are also a number of
particles travel from the connection point separating region to ``hot spot" and
get further acceleration, which is represented by the particles that travel a
large distance in z direction.

Figure 5 shows the same plot as Figure 4, except here the critical momentum is
$p_c = 10.0p_0$. It is shown again in Figure 5 $top$ that most of particles are
accelerated to high energy are in the hot spot. However, as opposite to Figure
4 $top$, there are very few particles accelerated at the center of plane since
energetic particles more transport away from the middle region and the time
available is not long enough. For a quick estimate, the acceleration time is
approximately,

\begin{eqnarray}
\tau_{acc} &= 
\frac{3}{U_1-U_2}\int^p_{p_0}(\frac{\kappa_{xx1}}{U_1}+\frac{\kappa_{xx2}}{U_2})
d\ln p \nonumber \\
&> \frac{3}{U_1-U_2}\int^{10p_0}_{p_0}(\frac{\kappa_\perp}{U_1}+\frac{\kappa_\perp}{U_2})
d\ln p > \tau_c \nonumber
\end{eqnarray}


Therefore for most of particles, they do not have sufficient time to be
accelerated to high energies at the center. A number of particles accelerated
at the center will travel to the ``hot spot" and get more acceleration, as
shown in Figure 5 $bottom$.

Clearly, this presents different pictures of particle acceleration by the shock
contain 2-D spatial magnetic field variations, indicates the resulting
distribution function is spatially dependent. In Figure 6 we show the steady
state energy spectra obtained in the regions \emph{top}: [$0.1<x<0.3$,
$\pi-0.2<z<\pi$] (black solid line), and [$0.1<x<0.3$, $-0.1<z<0.1$] (green
dashed line) and \emph{bottom}: [$0.8<x<1.0$, $\pi-0.2<z<\pi$] (black solid
line), and [$0.8<x<1.0$, $-0.1<z<0.1$] (green dashed line). It is shown that
the spatial difference among distribution functions at different locations
caused by large-scale magnetic field variation is considerable. The black lines
in both top plot and bottom plot, which correspond to the ``hot spots", show
power-law like distributions except at high energies. At high energies, the
particles will leave the simulation domain before gain enough energy which
causes the roll over in distribution function, this roll over is mainly caused
by a finite distance to upstream boundary. For other locations, the 2-D effect
we discussed will produce the modification in distribution functions. The most
pronounced effect can be found at the nose of the shock (green lines), in the
top panel the distribution of which shows a suppression of acceleration at all
the energies. This insufficient acceleration is most prominent in the range of
$6-12 p_0$. At these energies the acceleration time scales are longer than the
time for the field line convection swipe the particles away from the
connection-point separating region, as we discussed above. The bottom plot show
that deep downstream the spectrum of accelerated particles is similar at high
energies since the mobility of these particles.

\subsection{An oblique shock}
The previous discussion has established the effect of a spatially varying
upstream magnetic field on the acceleration of fast charged particles at a
shock which propagating normal to the average upstream magnetic field. We next
consider the case where the shock propagation direction is {\it not} normal to
the magnetic field.

Clearly, if the varying direction of the upstream magnetic field is such that
at some places the local angle of the magnetic field relative to the average
field direction exceeds the angle of the average magnetic field to the shock
plane, we will have situations similar to that discussed in the previous
sections. There will be places where the connection points of the magnetic
field to the shock move further apart or closer together. Hence we expect the
same physics can be applicable. An example is given in Figure 7. In this case
the ratio of $\delta B/B_0$ is taken to be $0.5$, the averaged shock normal
angle $\theta_{Bn} = 70^\circ$. It can be seen from this plot that the
connection points can still move toward each other in some regions. Figure 8
shows the density contours of accelerated particles the same as Figure 2, but
for the case of the oblique shock. We find that in this case the process we
discussed in the last section is still persistent, even for a oblique shock and
relative smaller $\delta B/B_0$. The ``hot spot" forms correspond to the
converging magnetic connection points and particle acceleration is suppressed
in the region where connection points separate from each other. We may conclude
that, for a shock which is oblique, if some magnetic field lines can intersect
the shock multiple times, we have ``hot spots" of accelerated particles forms
where the connection points converging together.

\section{Summary and Conclusions}

The acceleration of charged particles in shock waves is one of the most
important unsolved problems in space physics and astrophysics. The charged
particle transport in turbulent magnetic field and acceleration in shock region
are two inseparable problems. In this paper we illustrate the effect of a
large-scale sinusoidal magnetic field variation. This simple model allows a
detailed examination of the physical effects. As the magnetic field lines pass
through the shock, the connection points between field lines on the shock
surface will move accordingly. We find that the region where connection points
approaching each other will trap and preferentially accelerate particles to
high energies and form ``hot spots" along the shock surface, somewhat in
analogy to the ``hot spots" postulated by \citet{Jokipii2008AIP}. The shock
acceleration will be suppressed at places where the connection points move
apart each other. Some of the particles injected in those regions will
transport to the ``hot spots" and get further accelerated. The resulting
distribution function is highly spatial dependent at the energies we studied,
which could give a possible explanation to the \emph{Voyager} observation of
anomalous cosmic rays. Although we have discussed a simplified, illustrative
model, the resulting spectra and radial distributions show qualitative
similarity with the {\it in situ} {\it Voyager }observations. Thus, the
intensities do not in general, peak at the the shock and the energy spectra are
not power laws.  We show this process is robust even in the case of oblique
shocks with relatively small magnetic field variations. Large scale magnetic
field variation, which could be due to magnetic structures like magnetic
clouds, or the ubiquitous large scale field line random walk, will strongly
modify the simple planar shock solution. This effect could work in a number of
situations for large scale shock acceleration including magnetic variations,
for example, the solar wind termination shock and supernova blast waves.

\section*{Acknowledgement}
F. G. would like to thank Joe Giacalone for the tutorial and discussion of
stochastic integration method. F. G. also thank Chunsheng Pei and Erica McEvoy
for valuable discussion on stochastic differential equations. We acknowledge
partial support from NASA under grants NNX08AH55G and NNX08AQ14G.

\section*{Appendix: Study Diffusive Shock Acceleration using Stochastic Integration Method}

\noindent In a 2-D system we considered, the Parker transport equation $(1)$
can be re-written as,
\begin{eqnarray}
\frac{\partial f}{\partial t} &= \frac{\partial}{\partial
  x}\left(\kappa_{xx}\frac{\partial f}{\partial x} + \kappa_{xz}\frac{\partial f}{\partial z}\right)
  + \frac{\partial}{\partial
  z}\left(\kappa_{xz}\frac{\partial f}{\partial x} + \kappa_{zz}\frac{\partial f}{\partial z} \right) \nonumber \\
   &-U_{x}\frac{\partial f}{\partial x} +
  \frac{1}{3}\frac{\partial U_{x}}{\partial x} \left[\frac{\partial f}{\partial \ln p} \right] +
  Q
\end{eqnarray}

\noindent The equivalent stochastic differential equation is,
\begin{eqnarray}
\Delta x &=& r_1 (2\kappa_\perp \Delta t)^{1/2} + r_3
(2(\kappa_\parallel-\kappa_\perp)\Delta t)^{1/2}\frac{B_x}{B} \nonumber \\
&+& U_x\Delta t +
(\frac{\partial \kappa_{xx}}{\partial x} + \frac{\partial \kappa_{xz}}{\partial
z}) \Delta t \\
 \Delta z &=& r_2 (2\kappa_\perp \Delta t)^{1/2} + r_3
(2(\kappa_\parallel-\kappa_\perp)\Delta t)^{1/2}\frac{B_z}{B} \nonumber \\ 
&+& (\frac{\partial
\kappa_{zz}}{\partial z} + \frac{\partial \kappa_{xz}}{\partial x}) \Delta t \\
\Delta p &=& (1-\frac{1}{3} \frac{\partial U_x}{\partial x} \Delta t)
\end{eqnarray}

\noindent Where $r_1$, $r_2$, and $r_3$ are different sets of random numbers
which satisfy $<r_i> = 0$ and $<r_i^2> = 1$. In order to study diffusive shock
acceleration, we approximate the shock layer as a sharp variation $U_x =
(U_1+U_2)/2 - (U_1-U_2)tanh(x/th)/2$ with a thickness $th$ much smaller
compared with the characteristic length of diffusion acceleration
$\kappa_{xx1}/U_1$. At the same time, we have to make sure the time step
$\Delta t$ is small enough to resolve the motion in the shock layer.

\clearpage

\begin{figure}
\begin{center}
\hfill
\includegraphics[width=20pc]{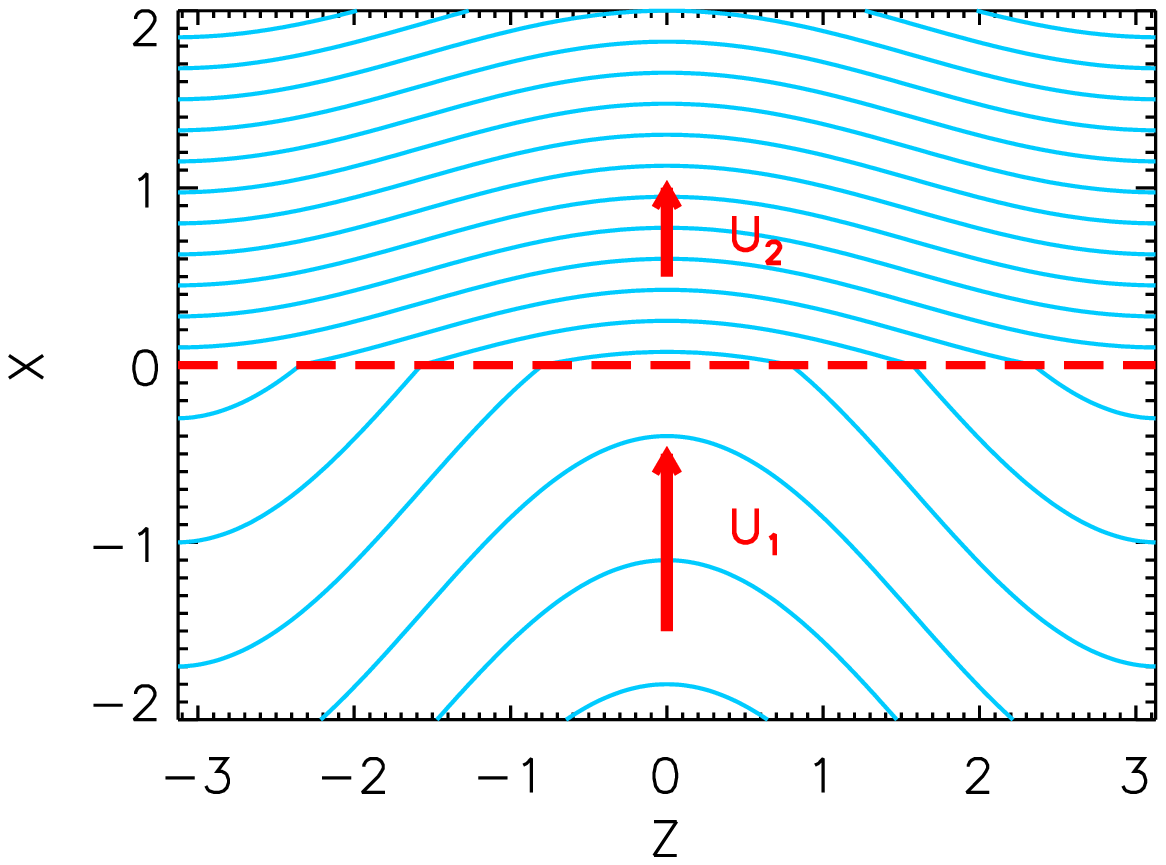}
\caption{The shock and the magnetic field geometry for an upstream average
magnetic field perpendicular to the shock normal. The blue lines
 represent the magnetic field lines and red dashed line indicates the surface of shock wave.
 The flow velocities are $U_1$ (upstream) and $U_2$ (downstream).}
 \end{center}
 \end{figure}

\begin{figure}
\begin{center}

\hfill
\includegraphics[width=25pc]{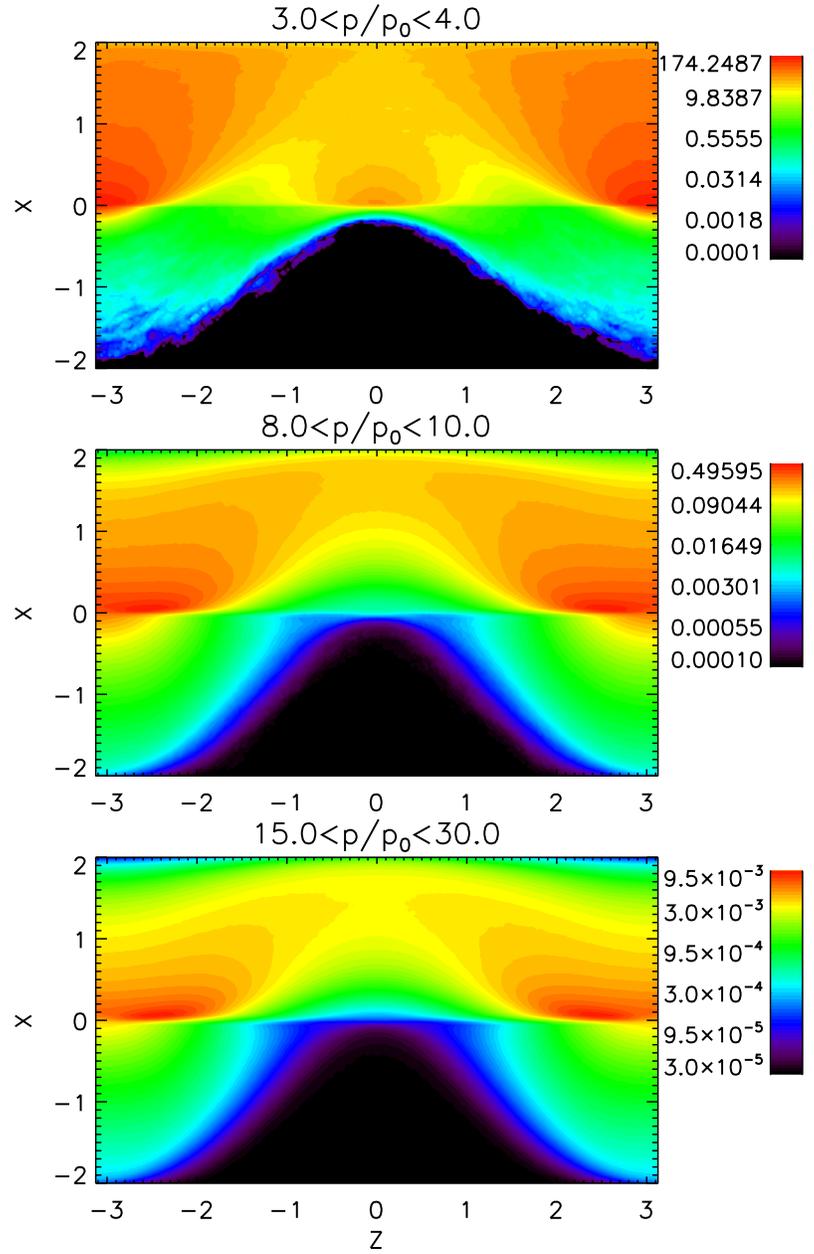}
\caption{The representation of density contour of accelerated particles, for
energy range: $3.0<p/p_0<4.0 (top)$, $8.0<p/p_0<10.0 (middle)$,
$15.0<p/p_0<30.0 (bottom)$. It is shown the hot spots forming on the both side
of the system. The acceleration at the center of the shock is suppressed.}
 \end{center}
 \end{figure}

\begin{figure}
\begin{center}
 \hfill
\includegraphics[width=25pc]{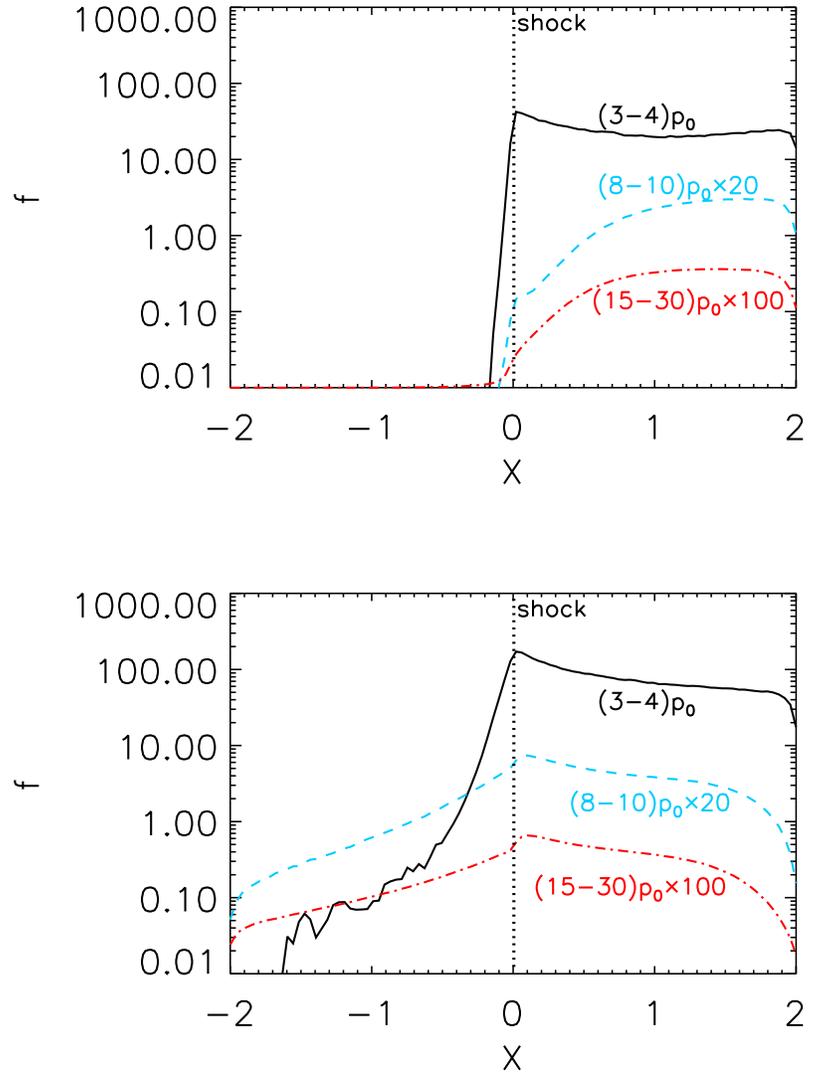}
\caption{The profiles of density of the accelerated particles, for energy
ranges: $3.0<p/p_0<4.0$(black lines), $8.0<p/p_0<10.0$(blue dashed lines),
$15.0<p/p_0<30.0$ (red dot dashed lines) at different locations $z = 0.0$ (top)
and $\pi$ (bottom), respectively. }
 \end{center}
 \end{figure}

\begin{figure}
\begin{center}
\hfill
\includegraphics[width=25pc]{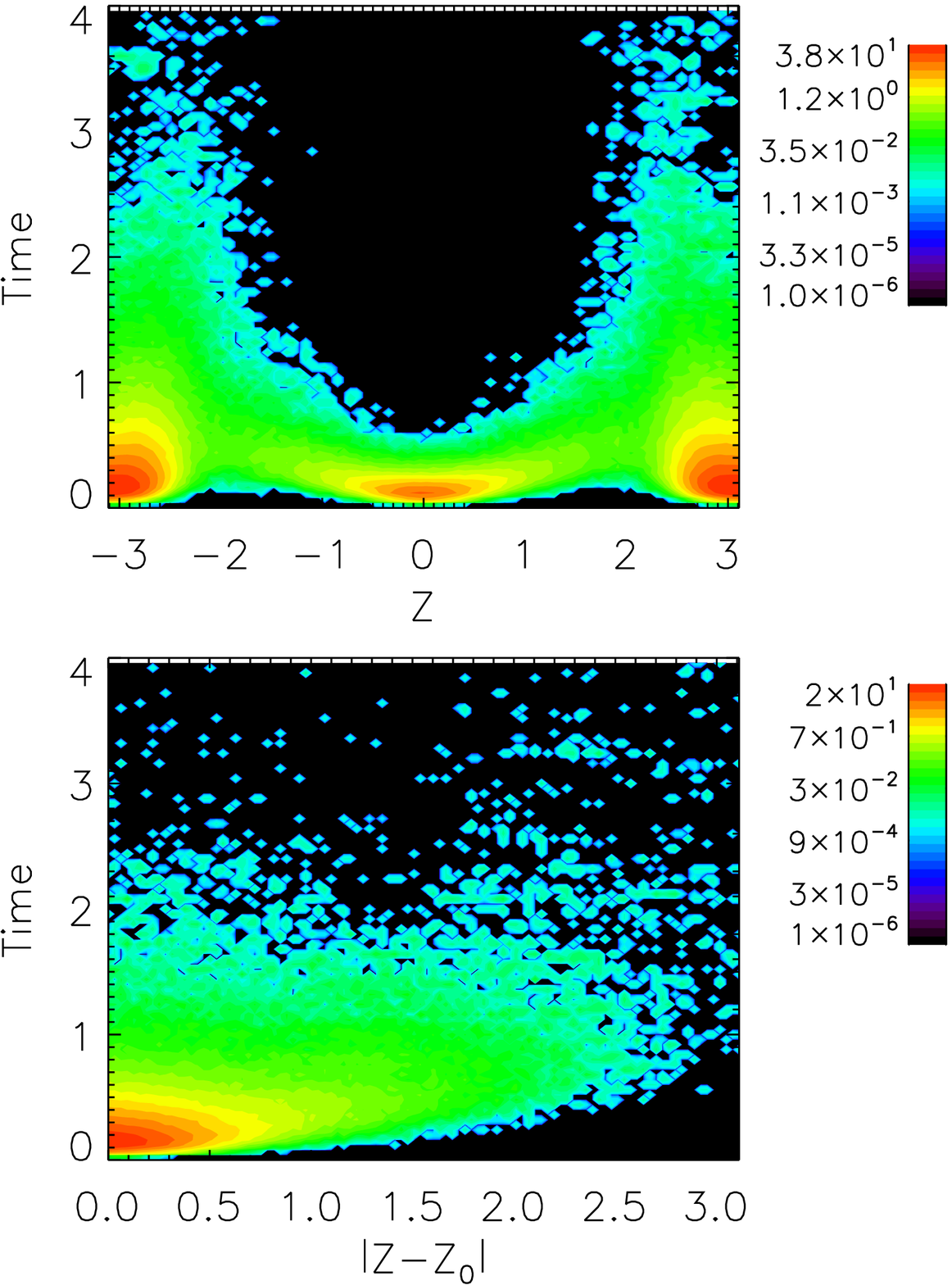}
\caption{$top$: The position in $z$ direction and time when the particle
momentum reached $p = 3.0p_0$; $bottom$: The travel distance in $z$ direction
$|z-z_0|$ and time when the particle momentum reached $p = 3.0p_0$.}
 \end{center}
 \end{figure}

\begin{figure}
\begin{center}
\hfill
\includegraphics[width=25pc]{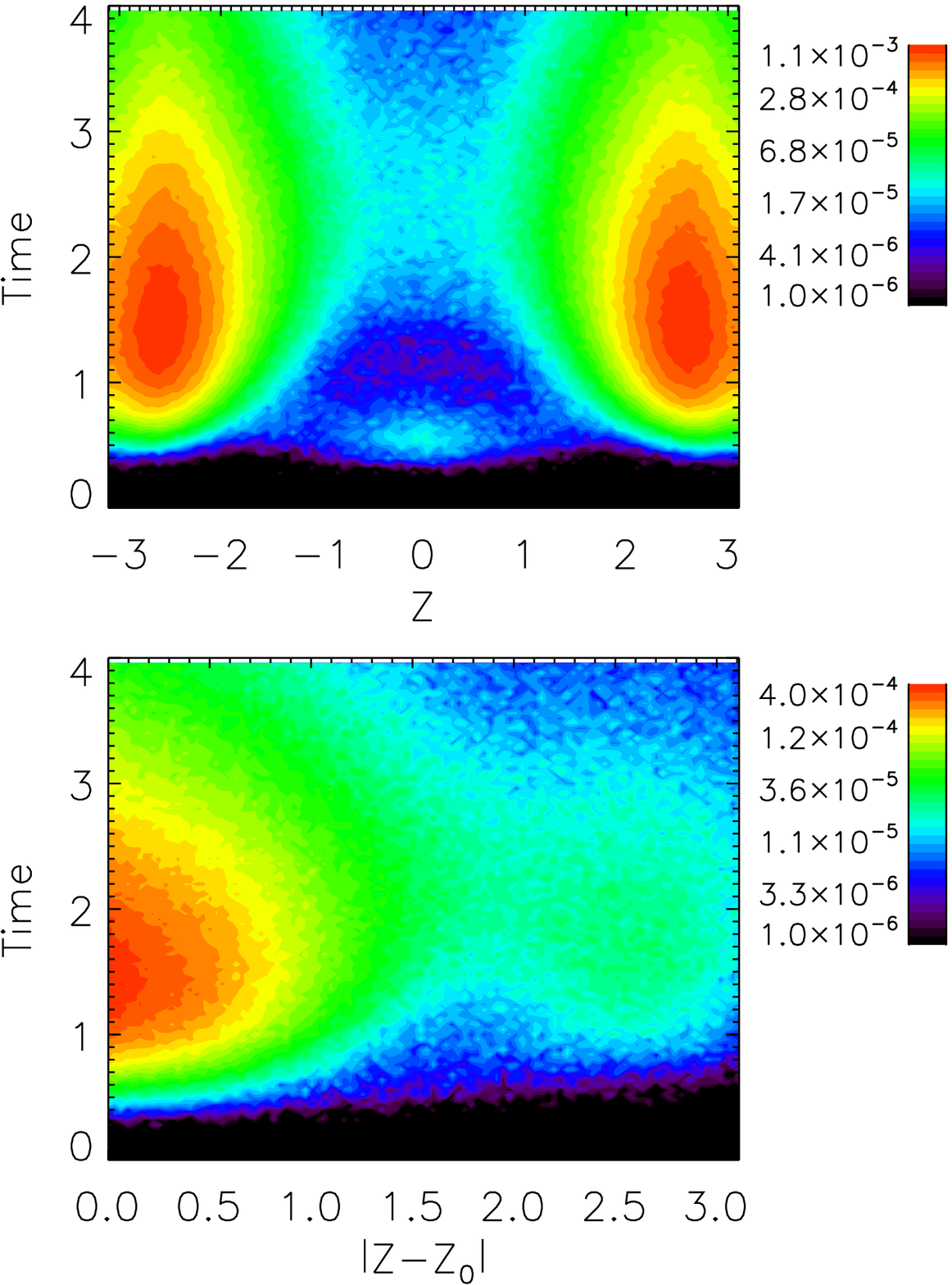}
\caption{$top$: The position in $z$ direction and time when the particle
momentum reached $p = 10.0p_0$; $bottom$: The travel distance in $z$ direction
$|z-z_0|$ and time when the particle momentum reached $p = 10.0p_0$.}
 \end{center}
 \end{figure}

\begin{figure}
\begin{center}
\includegraphics[width=25pc]{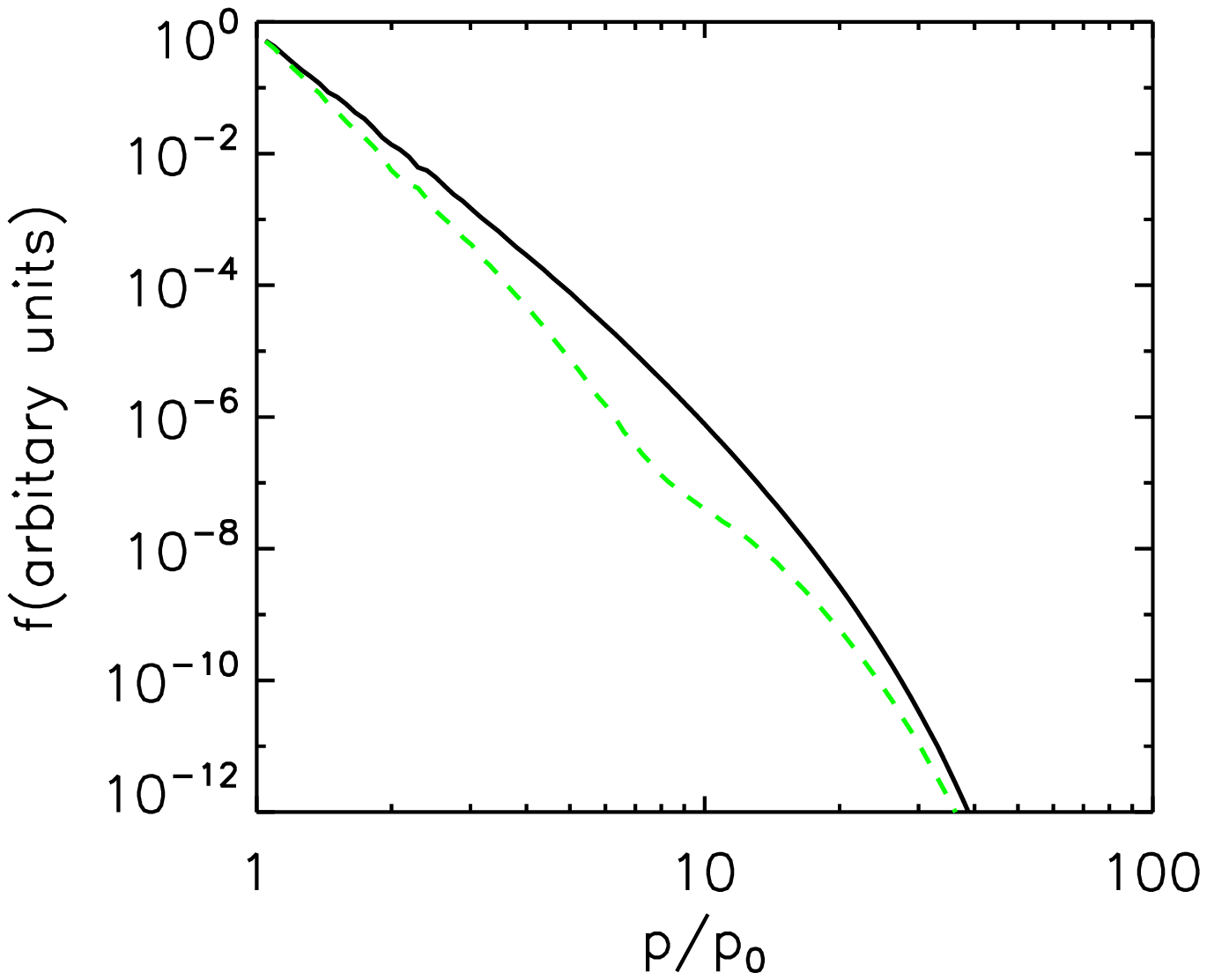}
\hfill
\includegraphics[width=25pc]{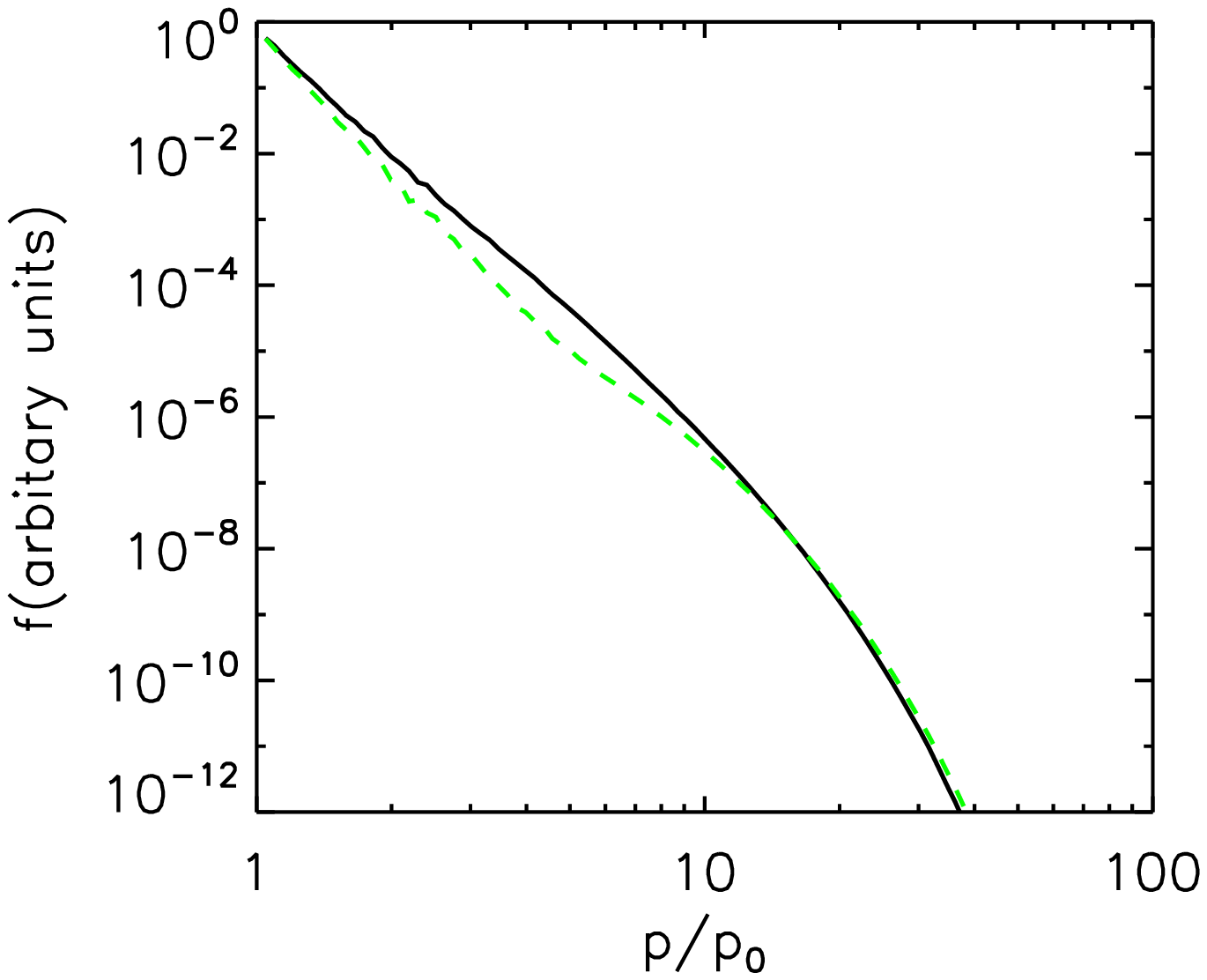}
 \caption{The steady state distribution functions at Top: [$0.1<x<0.3$,
$\pi-0.2<z<\pi$] (black solid line), [$0.1<x<0.3$, $-0.1<z<0.1$] (green dashed
line) and Bottom: [$0.8<x<1.0$, $\pi-0.2<z<\pi$] (black solid line),
[$0.8<x<1.0$, $-0.1<z<0.1$] (green dashed line), respectively.}
 \end{center}
 \end{figure}

\begin{figure}
\begin{center}
\hfill
\includegraphics[width=20pc]{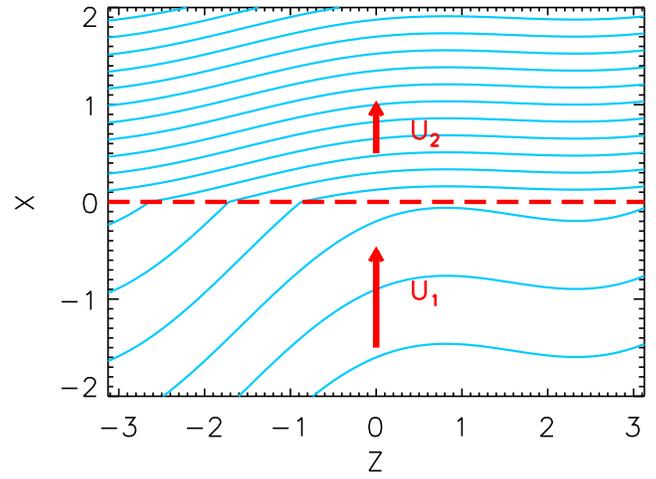}
\caption{The shock and the magnetic field geometry considered for the case of
an average magnetic field is $70^\circ$ of the shock normal. The blue lines
 represent the magnetic field lines and red dashed line indicates the position of shock wave.
 The flow velocities are $U_1$ (upstream) and $U_2$ (downstream).}
 \end{center}
 \end{figure}

\begin{figure}
\begin{center}
\hfill
\includegraphics[width=25pc]{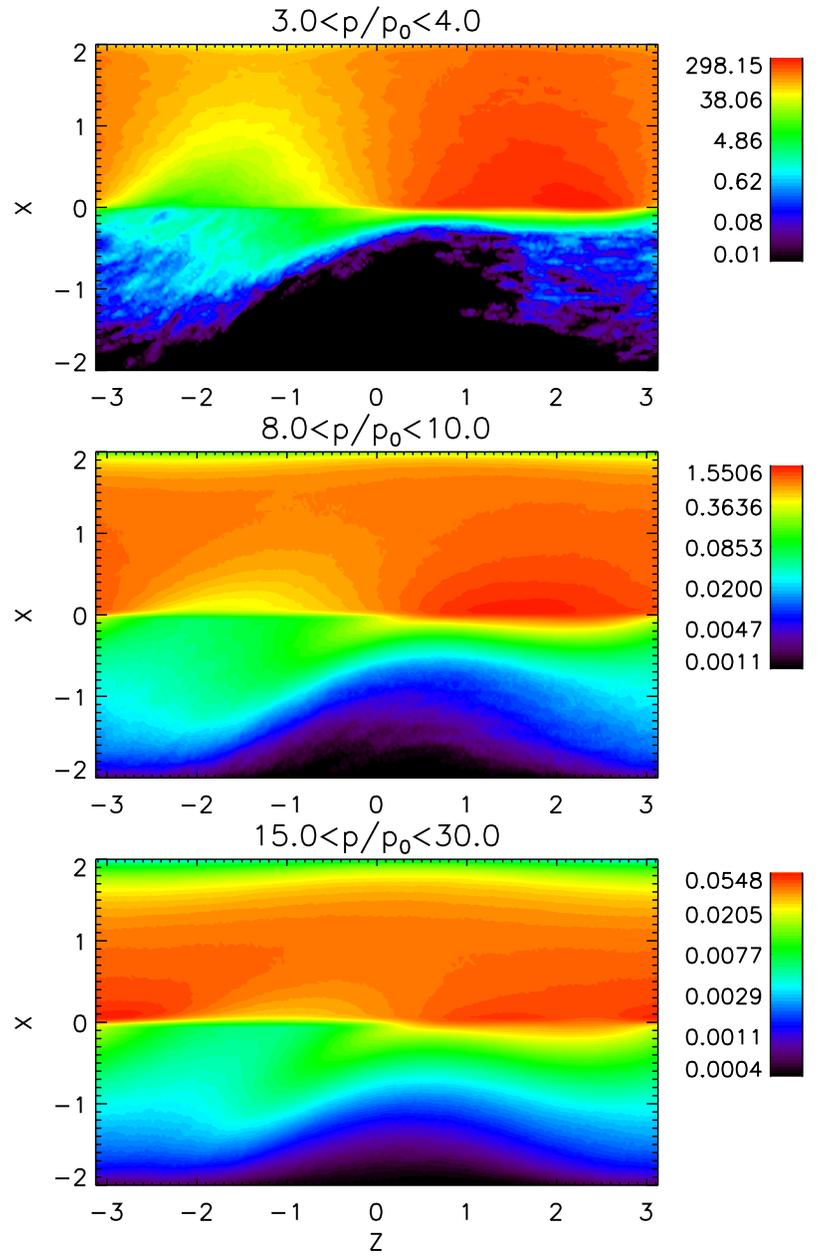}
\caption{The density contour of accelerated particles in the case of oblique
shock and $\delta B/B_0 = 0.5$, for energy range: $3.0<p/p_0<4.0 (top)$,
$8.0<p/p_0<10.0 (middle)$, $15.0<p/p_0<30.0 (bottom)$. }
 \end{center}
 \end{figure}

\end{document}